\documentclass[useAMS,usenatbib]{mnras}

\usepackage{stackengine} 
\usepackage{csquotes}

\def\apj{Astrophys. J.}  \def\aap{Astron. Astrophys.}
\def\mnras{Mon. Not. R. Astron. Soc.}  \def\prd{Phys. Rev. D}
  
\def\araa{Annu. Rev. Astron. Astrophys.}

\usepackage{tikz}
\usepackage{amsmath} \usepackage{amssymb} 

\begin{document}

\title{On the dynamics of tilted black hole-torus systems}
\author[V.~Mewes, F.~Galeazzi, J.~A.~Font, P.~J.~Montero and N.~Stergioulas]
{Vassilios Mewes$^{1}$\thanks{E-mail: vassilios.mewes@uv.es},
Filippo Galeazzi$^2$, 
Jos\'e A.~Font$^{1,3}$, 
Pedro J.~Montero$^{4}$, 
\and and 
Nikolaos Stergioulas$^5$ 
\\
  $^1$Departamento de Astronom\'{\i}a y Astrof\'{\i}sica,
  Universidad de Valencia, 46100 Burjassot (Valencia), Spain\\
  $^2$Lehrter Stra\ss e 51 HH, 10557 Berlin, Germany\\
  $^3$Observatori Astron\`omic, Universitat de Val\`encia, C/ Catedr\'atico 
  Jos\'e Beltr\'an 2, 46980, Paterna (Val\`encia), Spain \\
  $^4$Max-Planck-Institut f\"ur Astrophysik,
  Karl-Schwarzschild-Str.~1, 85741 Garching, Germany \\
  $^5$Department of Physics, Aristotle University of Thessaloniki,
  Thessaloniki 54124, Greece \\
  }
\date{\today}
\maketitle
\begin{abstract}
We present results from three-dimensional, numerical relativity simulations of a {\it tilted} black 
hole-thick accretion disc system. The simulations are analysed using tracer particles in the disc 
which are advected with the flow. Such tracers, which we employ in these new simulations for
the first time, provide a powerful means to analyse in detail the complex dynamics of 
tilted black hole-torus systems. We show how its use helps to gain insight in the overall
dynamics of the system, discussing the origin of the observed 
black hole precession and the development of a global non-axisymmetric $m=1$ mode 
in the disc. Our three-dimensional simulations show the presence of quasi-periodic 
oscillations (QPOs) in the instantaneous accretion rate, with frequencies in a range compatible with those
observed in low mass X-ray binaries with either a black hole or a neutron star component. The frequency 
ratio of the dominant low frequency peak and the first overtone is $o_1/f \sim 1.9$, a frequency ratio not 
attainable when modelling the QPOs as $p$-mode oscillations in axisymmetric tori.
\end{abstract}


\section{Introduction}
\label{sec:introduction}

The majority of accretion discs around Kerr black holes (BH) is 
believed to be tilted with respect to the equatorial plane of the
central BH (see~\citet{Fragile2001,Maccarone2002,Fragile2007a} 
for arguments). Recent fully general relativistic hydrodynamics (GRHD)
simulations of tilted black hole-neutron star (BHNS) mergers
have shown that a tilted, thick accretion disc can self-consistently
form in these events~\citep{Foucart2011,Foucart2013,Kawaguchi2015}.
In a tilted BH--torus system, the dynamics of the system is
fundamentally different from the aligned case due to 
general relativistic effects affecting inclined 
particle orbits in the Kerr spacetime,
such as the Lense--Thirring (LT) effect~\citep{Lense1918}.
The torque caused by the LT effect has a strong radial 
($r^{-3}$) dependence and causes the disc to start 
precessing differentially, as a result of which it might 
become {\it twisted} and {\it warped}, affecting its dynamical 
behaviour (see~\citet{Nelson2000} for a definition of twist and warp).
In thick, tilted discs around Kerr BHs, the evolution and propagation
of warps can be described by bending waves rather than 
diffusion~\citep{Ivanov1997,Demianski1997,Lubow2002}.
In particular, in these systems the tilt angle 
does not approach zero in the vicinity of the central 
BH, as one would expect if the viscous Bardeen-Petterson 
effect~\citep{Bardeen1975} would be at play. 
This behaviour of the radial tilt profile in the inner region of 
thick, tilted accretion discs has been observed in the inviscid 
GRHD simulations of~\citet{Fragile2005}, 
which were performed in a fixed background metric, as well as in our 
recent fully dynamical GRHD simulations~\citep{Mewes2016}, hereafter referred to as Paper I. 
Incorporating general relativistic effects in the simulations 
of these systems is necessary to obtain correct disc evolutions,
see for instance~\citet{Nealon2016}, where omitting or incorporating
general relativistic apsidal precession completely changes the
evolution of the tilt angle in the vicinity of the central BH.

In geometrically thick and radially slender discs such as the ones studied in
~\citet{Fragile2005} and Paper I, the response of the disc to the LT torque 
is solid body precession,
as the sound crossing time of the disc is much shorter than the 
LT timescale~\citep{Fragile2005}. The
precessing disc is exerting an equal and opposite torque on the 
central Kerr BH~\citep{King2005}, which should therefore start to 
precess as well, at least in those systems in which the disc mass 
is not negligible and the spacetime therefore cannot be assumed 
to be a fixed background. In
Paper I we indeed observed significant precession of the 
central Kerr BH for all models we studied.
As described by~\citet{King2005}, the LT torque alone does not act in a
direction that results in disc alignment. However, there have been early 
arguments that the central BH should eventually align with the
disc angular momentum~\citep{Rees1978,Scheuer1996}. In 
particular~\citet{King2005} argued that the total torque cannot have a
component in the direction of the BH spin, but can be broken
down to a contribution that induces precession and a second,
dissipative torque that tends to align the BH and the accretion disc.
Recently, in the post-merger phase of the BHNS merger simulations 
of~\citet{Kawaguchi2015}, the authors found a significant alignment of the 
accretion disc in a 
timescale compared to the accretion timescale of the disc. These authors
suspected that the alignment could have been due to the transport
of angular momentum induced by non-axisymmetric shock waves in
the disc. In Paper I, we also observed phases of significant 
alignment, particularly during the growth of a global, non-axisymmetric
instability in some of our disc models, which we identified as the
Papaloizou-Pringle instability (PPI)~\citep{Papaloizou1984}.

In this paper we revisit in detail one of the models we studied in Paper I,
model {\tt NC1a03b05}, performing two new simulations of a tilted black hole-torus
system, one with the spacetime fully evolved and a second one keeping 
the spacetime fixed (i.e.~in the so-called Cowling approximation). In these
new simulations we use fluid tracers placed in the disc at the beginning of the 
simulation that are advected with the flow. Such tracers provide an additional valuable
means to analyse the disc evolution. In performing these new simulations and analysing 
the disc evolution in greater detail, we gain insight on the origin of
the observed BH precession as well as in the development of the 
observed global non-axisymmetric $m=1$ mode in the disc and the resulting 
dynamics of the system. Non-axisymmetric $m=1$ modes are special in that, once they 
appear, the centre of mass of the disc does not coincide with the centre of mass of the 
system anymore, leading to a perturbed potential that causes a drift of the central
mass away from the centre of mass of the system
~\citep{Adams1989,Heemskerk1992}. The induced motion
of the central mass can also enhance the strength of the $m=1$ 
mode significantly~\citep{Adams1989,Christodoulou1992}.  

Furthermore, the growth of the $m=1$ mode modulates the
instantaneous accretion rate, a quantity which is commonly employed
as a measure of X-ray luminosity~\citep{Paradijs1988,Mendez1999}.
The presence of quasi-periodic oscillations (QPOs) in the accretion 
rate and their association with the disc evolution has been used to 
explain the QPOs observed in low mass X-ray binaries (LMXBs)~(see 
e.g.~\cite{Klis2006} and references therein). Here, we show that 
QPOs are indeed present in the accretion rate and that their 
frequencies are in a range compatible with those observed in LMXBs.

This paper is organised as follows:
We briefly describe the simulation details and the properties
of the initial model in Section~\ref{sec:numericalmethods}
Our results are presented and discussed in detail 
in Sections~\ref{sec:torques} and~\ref{sec:m1}. 
Finally, we present our conclusions in Section~\ref{sec:conclusions}. 
Unless specific units are given, throughout this paper we employ
geometrised units $c=G=M_{\odot}=1$ where
$c$, $G$ and $M_{\odot}$ are the speed of light, the gravitational constant and the 
solar mass, respectively. 

\section{Computational framework}
\label{sec:numericalmethods}

\begin{table*}
    \caption{\label{tab:initial-model} Main characteristics of
    the initial model {\tt NC1a03b05}. From left to right the columns indicate the central
    BH mass $M_{\mathrm{BH}}$,  the disc-to-BH mass ratio $q$, defined
    as the ratio of the total gravitational mass of the torus and the irreducible mass of the
    BH, the inner and outer
      disc radii $r_{\rm{in}}$ and $r_{\rm{out}}$, the maximum
      rest-mass density $\rho_{\mathrm{c}}$, the polytropic constant $K$ of the
      equation of state, the orbital period
      $P_{\mathrm{orb}}$ and orbital frequency $f_{\mathrm{orb}}$
      at the radius of the initial $\rho_{\mathrm{c}}$, the
      specific angular momentum profile in the equatorial plane of the
      disc $l$ (in terms of
      the Schwarzschild radial coordinate $R$), the BH spin parameter
      $a$, and the initial tilt angle $\beta_0$. The model is
      evolved with a $\Gamma$-law ideal fluid EOS with
      $\Gamma=4/3$.}
    \label{table:model}
      \begin{tabular}{ccccccccccc}
        \hline
        $M_{\mathrm{BH}}$  & $q$ & $r_{\rm{in}}$ & $r_{\rm{out}}$ & $\rho_{\mathrm{c}}$ & $K$ & $P_{\mathrm{orb}}$ & 
        $f_{\mathrm{orb}}$  & $l$-profile & $a$ & $\beta_0$
       \\
       & & & & & & [ms]& [Hz] & & &
       \\
       \hline
         $0.9775$ &  $0.110$ & 3.60 
         &  33.5 
         & $1.69 \times 10^{-5}$ 
         &$0.170$ 
          &  1.19 
         & 843 
          & 3.04 $R^{0.11}$ 
         & 0.3 
         & 5 \\
       \hline

      \end{tabular}
\end{table*}

The simulations reported in this paper use the same numerical relativity codes and techniques used for the 
simulations of Paper I. We briefly summarise here the basic ingredients, addressing the interested reader to Paper I
for details.  The simulations are performed with the Einstein Toolkit~\citep{Loffler2012}, an open source, numerical relativity
code that solves the Einstein equations in the $3+1$ formalism. The spacetime evolution is done with the 
{\verb McLachlan } code~\citep{Brown2009,Reisswig2011}, an implementation of the 
Baumgarte-Shapiro-Shibata-Nakamura (BSSN) formulation~\citep{Shibata1995,Baumgarte1998}, while
the hydrodynamics evolution is performed by the {\verb GRHydro } code~\citep{Baiotti2005,Loffler2012,Mosta2014}. 
The computational domain is composed of 13 nested mesh refinement
levels, provided by the {\tt Carpet} code~\citep{Schnetter-etal-03b}.
For the current simulations, the only technical difference lies in the evolution of the spacetime shift vector
$\beta^i$. While in the study performed in Paper I we used the dynamical 
$\eta$-damping described in~\citet{Alic2010}, in the current simulations
we use the standard $\Gamma$-driver
shift condition~\citep{Alcubierre2003} (see~\citet{Loffler2012} for the implementation details
in the Einstein Toolkit.) This is motivated by the greater compatibility with the
standard version of {\verb McLachlan } and also in order to 
check for the influence of the gauge evolution on the physics
we observed in Paper I.

The initial model for the current simulations is model {\tt NC1a03b05} of Paper I, which has a
non-constant specific angular momentum profile and
was shown to exhibit the development of a long-lasting 
non-axisymmetric $m=1$ mode. 
The parameters of this model are given in Table~\ref{table:model}.
We evolve the model up to 32 orbital periods in two different ways in order to gauge the influence of the 
self-gravity of the fluid in the evolution of this model. The first evolution takes into account the solution of
the coupled system of Einstein equations and GRHD equations, and thus corresponds to a fully
dynamical spacetime evolution. For the second evolution we assume the test-fluid (Cowling) approximation 
in which the spacetime is fixed and only the GRHD equations are solved. As initial data for this second simulation 
we use the data of the fully evolved spacetime run at 
$t=1.5 t_{\mathrm{orb}}$.
This is to ensure that we evolve the same system, and 
in particular to provide the same perturbation that triggers
the growth of the non-axisymmetric instability in both runs.

In order to compute the evolution of the tracer particles 
in our system, we adopt a simple test particle approximation to convert from the Eulerian 
representation of the fluid flow (in which the hydrodynamics variables are evolved on the computational grid) to the Lagrangian one, needed to compute the velocity of the particles. 
This method is often used in the context of core-collapse supernovae simulations to describe the evolution 
of the unbound matter subject to r-process nucleosynthesis (see \citet{Travaglio2004, Nakamura2015}).  
These passively advected particles allow us to record their velocity, internal energy and pressure by interpolating
the corresponding quantities from the underlying grid. We adopt a linear interpolation to project the physical 
quantities computed on the grid onto the tracer particles.

Using the local 3-velocity of the fluid, $\boldsymbol{v}$, lapse function
$\alpha$, and shift vector $\boldsymbol{\beta}$  
we can evolve the position of the particle by simply integrating 
\begin{align}
\frac{d\boldsymbol{x}}{ dt} = \tilde{\boldsymbol{v}}(\boldsymbol{x})\;, 
\end{align}
where $\tilde{\boldsymbol{v}} \equiv \alpha \boldsymbol{v} - \boldsymbol{\beta}$ is the advection
speed with respect to the coordinates~\citep{Foucart2014b}.
For the time evolution of the tracers, we use a second-order accurate in time Adams-Bashforth 
explicit integrator, which requires two previous time steps, $t^{n-1}$ and $t^{n}$, to compute the position vector, $\boldsymbol{x}$,
of the particles at $t^{n+1}$. 
As a result, we can express the evolution equation for the position as
\begin{align}
\boldsymbol{x}^{n+1} = \boldsymbol{x}^{n} + \frac{3}{2} \Delta t 
\left(\tilde{\boldsymbol{v}}^{n}-\tilde{\boldsymbol{v}}^{n-1})\right)\;,
\end{align}
where $\Delta t$ represents the time step relative to 
the finest mesh refinement where the tracer particle is located.

Initially we place 250000 tracers at random locations according to 
the underlying rest-mass density $\rho$ distribution in the accretion disc, 
resulting in all tracer particles being of equal mass ($3.97\times
10^{-7}$)  initially and during the
evolution. The bulk motion of the tracers is able to capture accurately the motion
of the underlying fluid flow especially when the fluid flow is not turbulent and 
strong shocks are not present.

Using this prescription, the tracers are then advected with the fluid flow 
during the evolution of the disc. The tracers are output in {\tt hdf5} snapshots at 
user-specified time intervals during the evolution, which allows one to perform 
the disc analysis as a post-processing step rather than during the simulation. 
The analysis of the disc via the tracer particles performs very 
well in the bulk regions of the disc and accurately reflect the 
total mass, energy and angular momentum of the disc during the
evolution. For the analysis of complicated flow details in low density 
regions (such as the accretion streams observed 
in~\citet{Fragile2005} and in Paper I), however, 
the disc analysis thorn described in Paper I performs much more accurate as
it analyses the disc morphology using the full 3D data during the evolution.
Due to the much larger output files resulting from the output 
of the 3D spacetime and hydrodynamics variables, the disc analysis
has to be performed during the simulation rather than as a 
post-processing step.
Depending on the number of tracers, it might be computationally cheaper 
to advect the tracer particles with the flow and later 
analyse them rather than performing the complicated disc analysis
during the simulation. This may significantly reduce the runtime of the evolution 
simulations. Due to the manageable file sizes of the tracer output, 
the tracers are furthermore a very powerful tool for the visualisation of 
the disc evolution.

\section{Results}
\label{sec:results}

\subsection{Black hole precession from Lense-Thirring torque}
\label{sec:torques}

\begin{figure}
  \centering
  \includegraphics[scale=0.9]{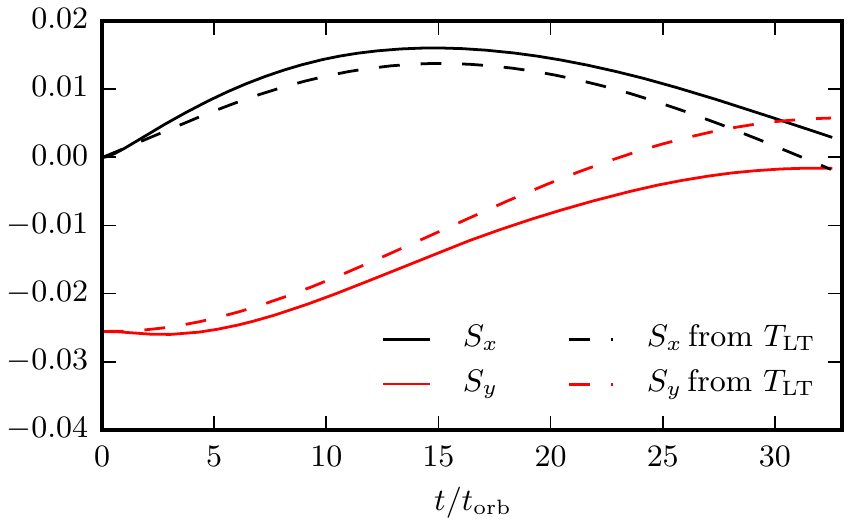}
  \caption{Evolution of the $x$ and $y$ components of the BH spin
    $\boldsymbol{S}$ and the evolution of the spin components
    resulting from the LT torque exerted by the disc. }
  \label{fig:LT_torque}
\end{figure}

The main driver of the tilted disc evolution is the Lense-Thirring 
(LT) precession. At the lowest post-Newtonian order of spin-orbit interactions,
a particle in an inclined orbit around a central Kerr BH
will be subject to the following torque~\citep{Kidder1995,Merritt2012}: 
\begin{eqnarray}\label{eq::LT_prec}
\boldsymbol{T}_{\mathrm{LT}} = 2 \frac{\boldsymbol{S} \times \boldsymbol{J}}{r^3},
\end{eqnarray}
where $\boldsymbol{S}$ is the BH spin, $\boldsymbol{J}$ is the
angular momentum vector of the particle, and $r$ is the distance of the particle from the 
BH centre. This is the LT torque exerted by the central
BH on the particle. As a result of this torque, the orbital angular
momentum of the particle will start to precess around the BH spin. As there
is a strong radial dependence on the magnitude of the torque,
the inner regions of a tilted accretion disc will become twisted, i.e.~they will
differentially precess around the central BH. 
However, as shown in the results of~\citet{Fragile2005} and in Paper I, the 
sound crossing time in the disc is shorter than the LT timescale, 
which results in a global disc response
to the LT torque produced by the Kerr BH, leading to solid body
precession of the disc. The
disc exerts a torque of equal magnitude and opposite direction
on the BH, causing it to precess as well. 

\begin{figure}
  \centering
  \includegraphics[scale=0.9]{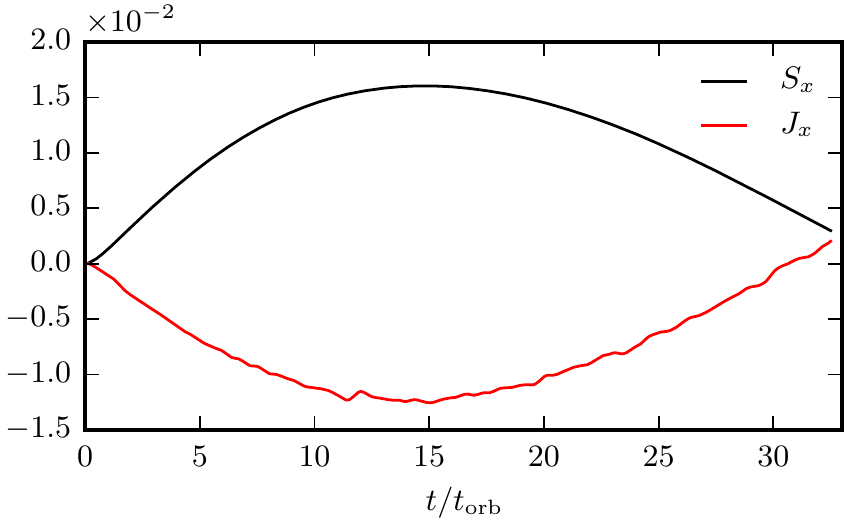}
  \\
  \vspace{-0.5cm}
  \includegraphics[scale=0.9]{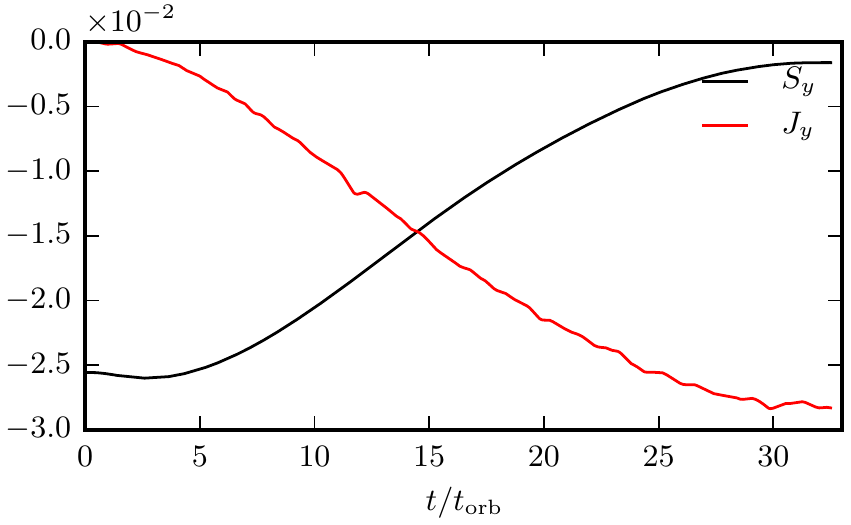}
  \caption{Evolution of the $x$ and $y$ components of the BH spin
    $\boldsymbol{S}$ and total disc angular momentum $\boldsymbol{J}_{\mathrm{disc}}$. }
  \label{fig:spin_J_components}
\end{figure}

To check whether the precession
of the BH we observe in our simulations has a physical origin, we compare the
time evolution of the $x$ and $y$ components of the BH spin, $S_x, S_y$, with
the evolution of the BH spin that would have resulted from summing up
the cumulative 
spin change resulting from the LT torque of the disc at each timestep:
\begin{equation}
\boldsymbol{S}_{\mathrm{LT}}(t) = \boldsymbol{S}(0) 
+ \int_0^t \boldsymbol{T}_{\mathrm{LT}} dt.
\end{equation} 
As the disc is approximately precessing as a solid body, we calculate
the torque coming from the disc total angular momentum vector, 
$\boldsymbol{J}_{\mathrm{disc}}$,
choosing a radius of $r=15.0$ in Eq.~(\ref{eq::LT_prec}).
The results are shown in Figure~\ref{fig:LT_torque}. This figure
shows that the time evolution of the BH spin components
agree well with the torque exerted on the BH by the disc. 
In Fig.~\ref{fig:spin_J_components} we show the 
time evolution of the $x$ and $y$ components of $\boldsymbol{S}$
and $\boldsymbol{J}_{\mathrm{disc}}$. Clearly,
the sum of the spin components and disc angular momentum components
along the two directions is seen to be approximately constant for the
evolution, in agreement with the fact that the LT torque is the main 
driver for both disc and BH precession. 

\begin{figure*}
  \centering
  \includegraphics[scale=1.25]{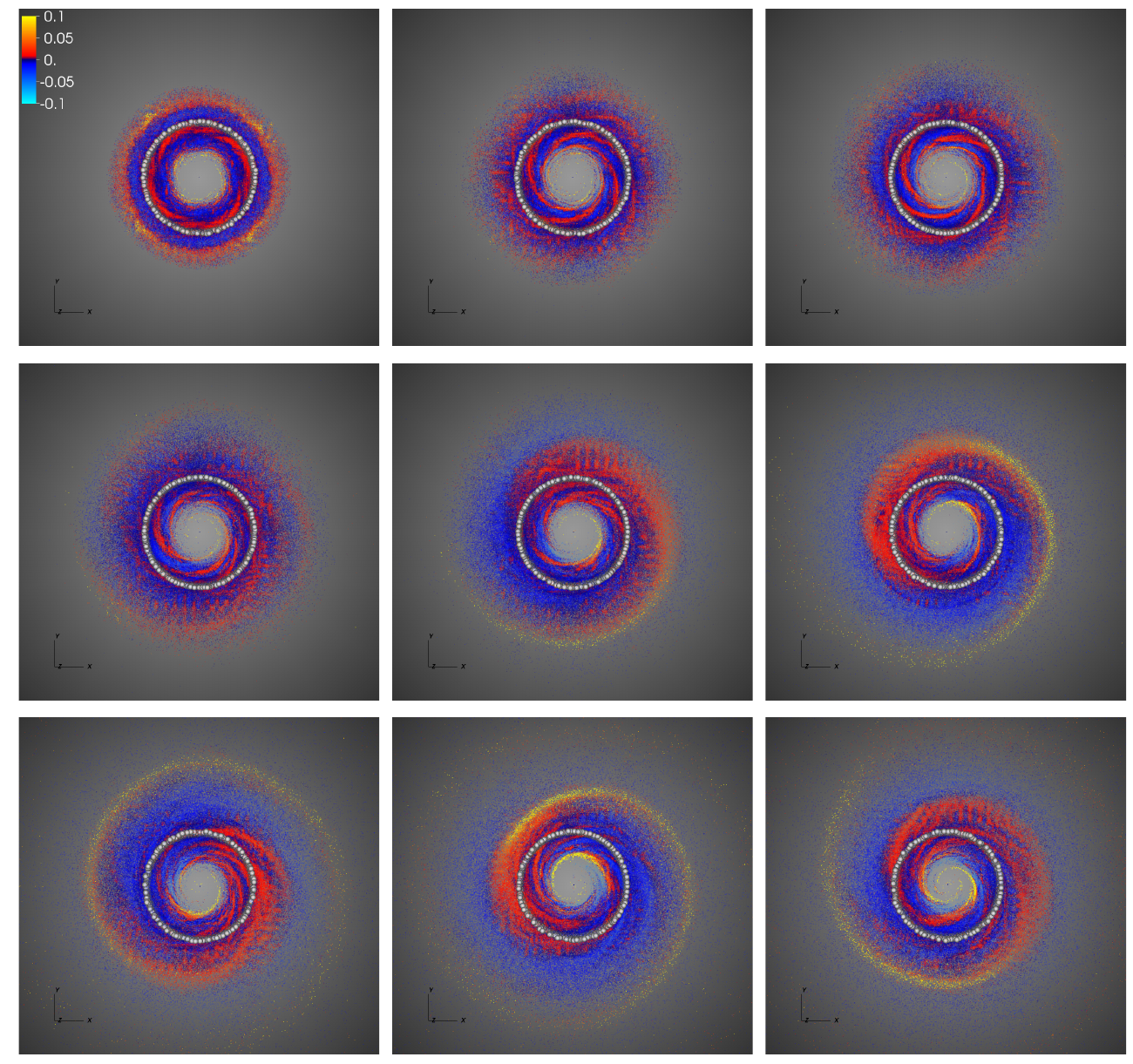}
   \caption{Fractional change in the rest mass density $\Delta \rho/\rho$ between two 
   timesteps, shown at 9 different snapshots of the evolution, 
   $t/t_{\rm orb} = 1, 4, 8, 12, 16, 20, 24, 28$, and 32 (from the top-left panel). 
   The domain shown in all panels is 100 ($\sim 148$ km) across both axes. 
   The grey circles show the tracers at the location of the corotation radius 
   $r_{\rm co}$. The formation of a spiral density wave from the growth of 
   the $m = 1$ mode is noticeable.} \label{fig:drho_rho_solo}
\end{figure*}

\subsection{$m=1$ non-axisymmetric instability}
\label{sec:m1}

\subsubsection{Spiral density wave}
\label{sec:spiral_density_wave}

\begin{figure}
  \centering
  \includegraphics[width=0.35\textwidth]{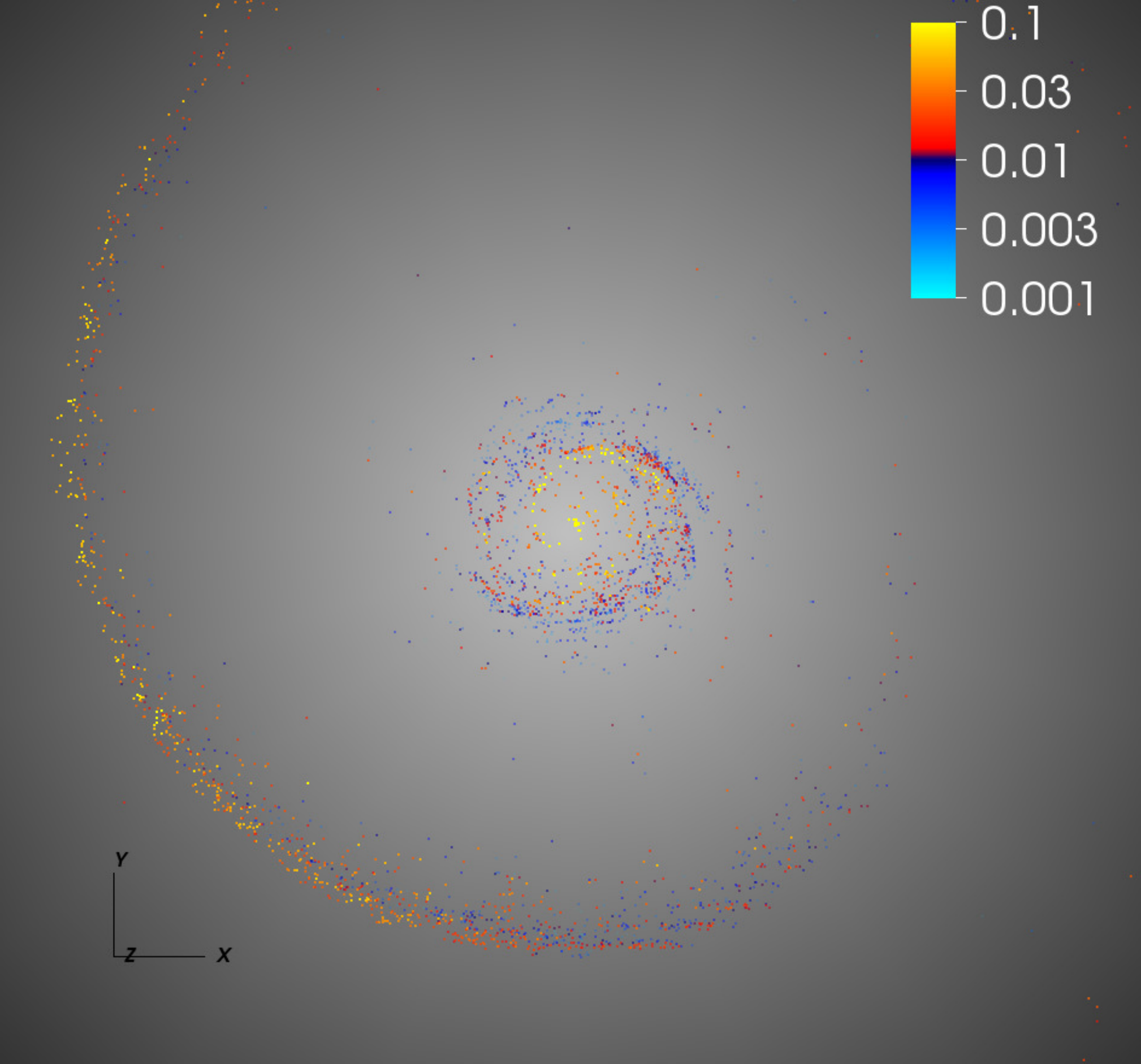}
  \caption{Fractional change in the fluid entropy $S$
     between two timesteps, for tracers with $\Delta \rho/\rho > 0.1$, 
     shown at $t/t_{\rm{orb}}=32$.}
  \label{fig:dS_S_drho_rho}
\end{figure} 

Two of the three models we studied in Paper I were found
to develop a global non-axisymmetric instability with $m=1$
being the dominant azimuthal mode. One such instability is the
instability of tori with constant specific angular momentum 
profiles, the Papaloizou-Pringle instability~\citep{Papaloizou1984}.
Early numerical investigations of~\citet{Hawley1987} showed that
the non-linear regime of the PPI resulted in the formation of
counterrotating overdensity lumps, which were dubbed `planets'.
The precise mechanism of the instability was elucidated 
by~\citet{Goodman1987} who showed that the planets
found in~\citet{Hawley1987} might be a new equilibrium configuration
of the fluid. The instability was subsequently
investigated numerically by~\citet{Hawley1991} as
well as in the recent, fully general relativistic simulations
of~\citet{Korobkin2011,Kiuchi2011} and those presented in
Paper I. 

We note here that the role played by magnetic fields in connection with 
non-axisymmetric hydrodynamical instabilities such as the PPI
is not yet fully understood. The interest in the PPI quickly diminished
after the discovery that the magneto-rotational 
instability (MRI)~\citep{Velikhov1959,Chandrasekhar1960} is
active in accretion discs~\citep{Balbus1991}, because accretion has been
seen to effectively stabilise a disc against the development of the 
PPI~\citep{Blaes1987}. However,~\citet{Fu2011} suggest that highly 
magnetised tori might still be unstable to global hydrodynamic instabilities. 
We plan on studying the effects played by magnetic fields on the 
development and growth of global non-axisymmetric instabilities 
through numerical relativity simulations of magnetised accretion tori 
around BHs in future work.

In this section we show that the growth of the $m=1$ mode
in the disc results in the formation of a spiral density wave with
a constant global pattern speed $\Omega_P$. As the disc is
differentially rotating this means that there is a location in the
disc, the so-called corotation radius $r_{\rm co}$, where the spiral
density wave has the same orbital velocity as the fluid in the
disc. Inside $r_{\rm co}$ the wave travels slower than the fluid, which
means that it has negative angular momentum with respect to the
fluid. While the wave amplitude is linear, it does not interact 
with the fluid, but once non-linear amplitude effects come
into play, the spiral density wave can couple to the fluid via
dissipation~\citep{Papaloizou1995,Goodman2001,Heinemann2012}. 
When this happens, the wave begins to transport angular momentum 
outwards as the fluid loses angular momentum to the wave inside the 
corotation radius. 
The development and persistence of the spiral density wave is shown 
in Figure~\ref{fig:drho_rho_solo}. The figure shows the fractional
change in density between two successive snapshots of the tracer
particles $\Delta \rho/\rho$ at 9 different times of the disc evolution, 
as well as the tracer particles that
are located at $r_{\rm co}$ (grey circles). 
The location of the corotation radius is defined as the radius where 
the $m=1$ pattern speed and the fluid orbital velocity are equal.
As usual, the non-axisymmetric modes in the disc are analysed
by means of an azimuthal Fourier transform of the rest-mass
density $\rho$~\citep{Zurek1986,Heemskerk1992}:
\begin{equation}
  D_m = \int \, \alpha \, \sqrt{\gamma} \,\rho \, e^{-i\,m\,\phi} \, d^3x \, ,
  \label{eq:mode-calculation}
\end{equation}
with $\alpha$ and $\gamma$ being the spacetime lapse function and the determinant of the
3-dimensional metric, respectively.
Similarly, the mode analysis performed using the tracers is 
accomplished by means of the following sum
\begin{equation}
  D_m = \sum_j^N \, m_j \, e^{-i\,m\,\phi} ,
  \label{eq:mode-calculation-tracers}
\end{equation}
where $m_j$ is the mass of each tracer particle.
From the mode amplitudes, the pattern speed of an azimuthal mode 
with mode number $m$ is defined as (see, for instance~\citet{Heemskerk1992})
\begin{equation}
\Omega_{P} = \frac{1}{m}\frac{d \phi_m}{dt},
\end{equation} 
where the phase angle $\phi_m$ is given by
\begin{equation}
\phi_m = \mathrm{tan}^{-1} \left(\frac{\mathrm{Im}(-D_m)}{\mathrm{Re}(D_m)} \right).
\end{equation}
For our simulation, we obtain an orbital period of the $m=1$ pattern
of $P_{P} \sim 1.96$ ms, which is slightly shorter than twice the initial 
orbital period of the disc at the location of the density maximum 
(see Table~\ref{table:model}).
We note that we have chosen the same range of the fractional change in the rest-mass density in 
all snapshots shown in Figure~\ref{fig:drho_rho_solo}, which corresponds to the interval
$[-0.1,0.1]$. At these wave amplitudes, non-linear effects are negligible compared to the linear
effects~\citep{Masset1997}. This plot range has however been chosen
for visualisation purposes and the wave amplitudes are actually 
much larger than $0.1$ after the saturation of the 
$m=1$ growth, reaching almost unity. At these wave amplitudes
non-linear effects are important and the spiral density wave can
couple to the fluid via dissipation.
To show the development of weak shocks, we plot the 
fractional change in the entropy $S=p/\rho^{\Gamma}$ for
tracers with $\Delta \rho/ \rho > 0.1$ 
in Figure~\ref{fig:dS_S_drho_rho}. From this plot, we clearly see
an increase in the fluid entropy in the inner and outer 
regions of the spiral density wave. 
In some snapshots of Figure~\ref{fig:drho_rho_solo}, we
can see fractional changes in $\rho$ in the form of two
spiral arms in the inner regions of the disc. The dominant $m=1$ 
mode developing in the disc should not produce these, which means
there could be a different mechanism at play. Standing shocks in the
inner regions of tilted accretion discs in the form of two 
spiral arms have been observed and analysed in~\citet{Henisey2012,Generozov2014}
and are a distinctive feature of tilted accretion discs.
Finally, we note that the mesh refinement boundaries of our computational grid also
produce entropy changes due to numerical dissipation. However,
these changes are about an order of magnitude smaller 
than the physical increase of entropy due to the 
development of shocks in the spiral density wave.

\subsubsection{Gravitational torque and motion of the central black hole}
\label{sec:bh_movement}

\begin{figure}
  \centering
  \includegraphics[scale=0.9]{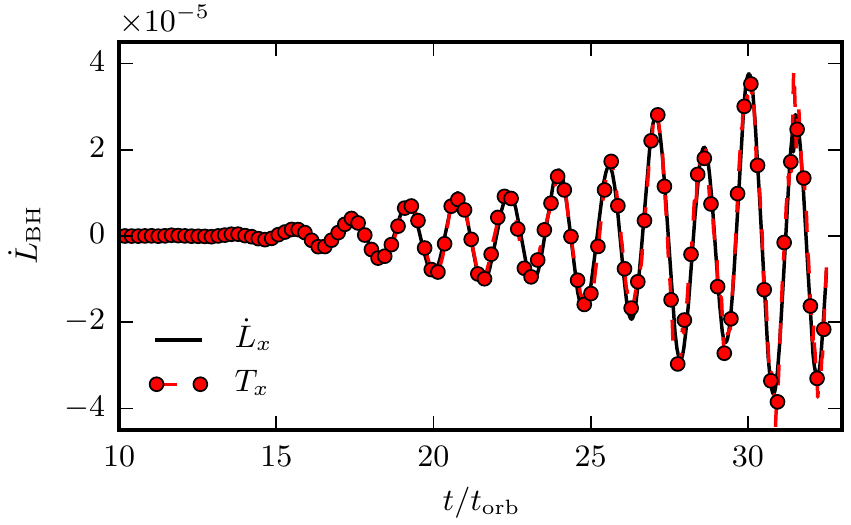}
  \\
  \vspace{-0.5cm}
  \includegraphics[scale=0.9]{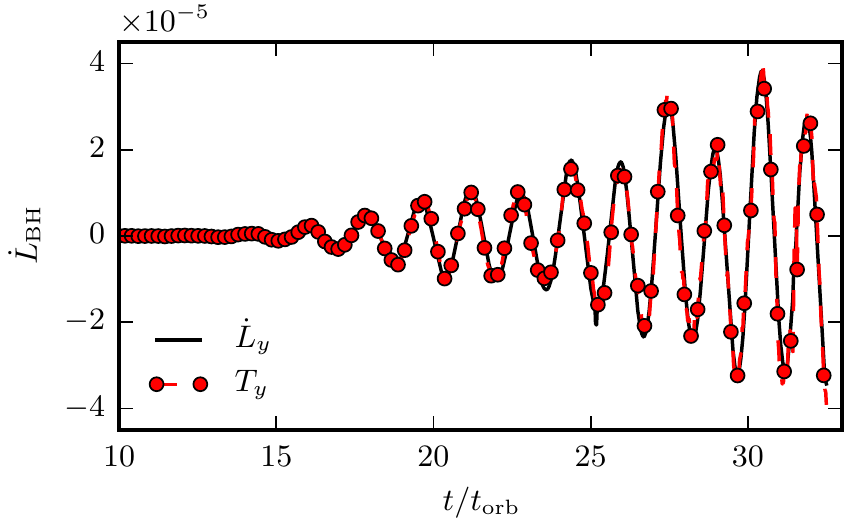}
  \\
  \vspace{-0.5cm}
  \includegraphics[scale=0.9]{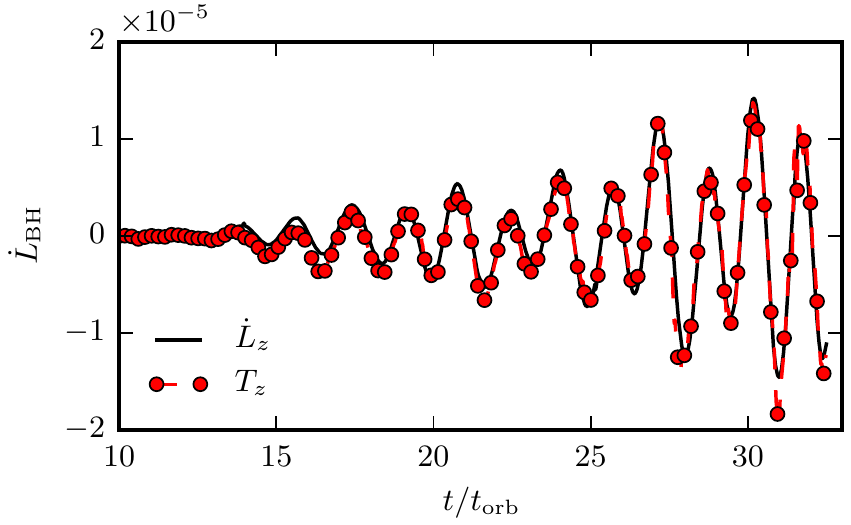}
  \caption{Time derivative of the components of the BH orbital angular
    momentum vector $\boldsymbol{L}_{\mathrm{BH}}$ and the components of the
gravitational torque computed from the non-axisymmetric matter distribution $\boldsymbol{T}_{\mathrm{G}}$. }
  \label{fig:gravitational_torque}
\end{figure} 

We next turn to the interaction of the non-axisymmetric
$m=1$ mode and the central BH. The moving overdensity
planet in the disc represents a time-changing non-axisymmetric
gravitational potential. 
Following~\citet{Roedig2012}, we estimate
the total gravitational torque on the BH exerted by the fluid is 
estimated by using the tracer particles at time $t$ in the 
following way:
\begin{equation}
\boldsymbol{T}_{G}(t) = \sum_{i}^{N} M_{\mathrm{BH}}(t)\, m_i\,
\frac{\boldsymbol{r}_{\mathrm{BH}}(t) \times (\boldsymbol{r}_i(t) -
  \boldsymbol{r}_{\mathrm{BH}}(t))}
{|\boldsymbol{r}_i(t) -
  \boldsymbol{r}_{\mathrm{BH}}(t)|^3},
\end{equation}
where the sum runs over all tracer particles and the time
dependent BH mass $M_{\mathrm{BH}}(t)$ 
and BH position vector $\boldsymbol{r}_{\mathrm{BH}}(t)$
are obtained using the 
{\tt QuasiLocalMeasures}~\citep{Dreyer2003,Schnetter2006} and 
{\tt AHFinderDirect}~\citep{Thornburg2003} thorns, respectively.

The resulting components of the gravitational torque
$\boldsymbol{T}_{\mathrm{G}}$ and the time derivative of the 
components of the BH orbital angular momentum 
$\boldsymbol{\dot{L}}_{\mathrm{BH}}$, are shown in 
Figure~\ref{fig:gravitational_torque}. This figure clearly shows
that the orbital angular momentum of the BH is caused
by the gravitational torque of the non-axisymmetric
matter distribution in the disc.
The orbital angular momentum components of the BH are calculated 
using the flat space coordinate rotational Killing vectors on the 
apparent horizon (AH)~\citep{Campanelli2007} (which gives the
Komar angular momentum in axisymmetry~\citep{Mewes2015})
without the subtraction of the position of the BH centre and
the subsequent subtraction of the intrinsic spin
of the AH. While this is by no means a gauge
invariant measure of the orbital angular momentum of the BH,
it agrees very well with the Newtonian calculation of the BH orbital
angular momentum as 
\begin{equation}
\boldsymbol{L}_{\mathrm{BH}} = M_{\mathrm{BH}} \, \boldsymbol{r}_{\mathrm{BH}} 
\times \boldsymbol{v}_{\mathrm{BH}}.
\end{equation}
The disc acquires some bulk orbital angular momentum (equal and 
opposite to that of the BH) as a result of the growth and the 
longevity of the $m=1$ mode as well. During the later stages of the
evolution, after the saturation of the $m=1$ mode growth, 
this bulk orbital angular momentum of the disc becomes comparable to the 
fluid angular momentum in the $x$ and $y$ directions.
It is precisely the evolution of the bulk orbital angular momentum that we
mistakenly identified as tilt and twist oscillations in Paper I.
This was due to an error in the reading of the location of the 
BH centre in the disc analysis code described in Paper I, which ultimately resulted 
in the calculation of the disc angular momentum about the grid origin
rather than about the BH centre.

As a result of this error in the disc analysis used for the results
presented in Paper I, we speculated that the 
observed global oscillations in the tilt
and twist might be due to the development of the 
Kozai-Lidov (KL) effect~\citep{Kozai1962,Lidov1962}.
This effect results in the periodic exchange of
the inclination and eccentricity of a particle orbit around a
central mass that is itself in an orbit with a third mass (the so-called 
perturber). The effect is a direct consequence of the
conservation of the total angular momentum of the system.
As the KL effect has recently been observed to operate in inclined hydrodynamical
discs around a central mass which is in a binary with another 
mass~\citep{Martin2014}, we assumed in Paper I that the KL effect 
might be in operation in our discs as the long-lasting $m=1$ mode
forces the central BH to move in a quasi-binary orbit with
the overdensity planet in the disc. To demonstrate the operation of
the KL effect and to explain the oscillations in more detail has been
one of the main motivations for carrying out new simulations and
employing the tracer particles as a new tool for the analysis of the 
disc. As explained above, we have clarified the origin of the global
tilt and twist oscillations as being connected to the bulk orbital 
angular momentum the disc acquires when forming the 
quasi-binary system with the BH. The KL effect is therefore
not seen to be in operation in our discs.

\begin{figure}
  \centering
  \includegraphics[scale=0.9]{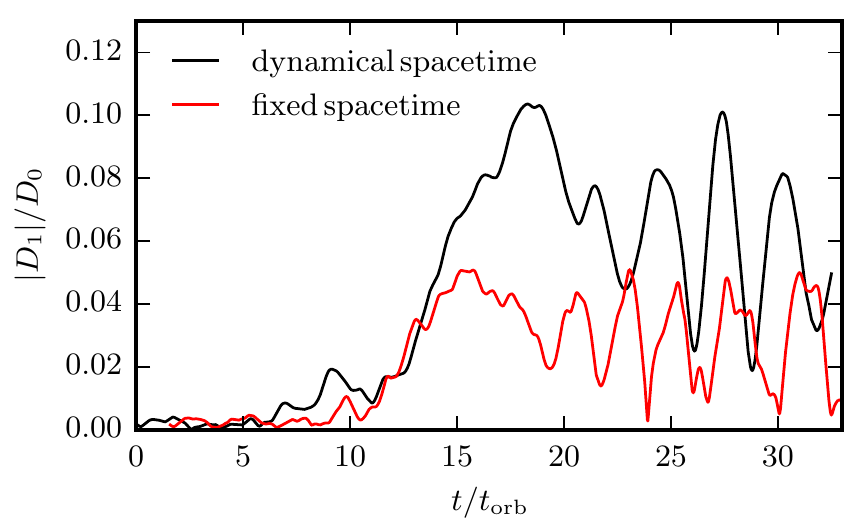}
  \caption{Evolution of the $m=1$ mode for the dynamical spacetime simulation (black line) and for the
    fixed spacetime simulation (red curve), calculated with the tracer
    particles.}
  \label{fig:PPI_growth_rate}
\end{figure} 

The effect of the motion of the central BH on the strength
of the $m=1$ mode is shown in Figure~\ref{fig:PPI_growth_rate}. This
figure displays the time evolution of the normalised magnitude of the $D_1$ mode for
our two simulations. 
While growth rates of the $m=1$ mode strength are very similar
for the two runs, the maximum mode amplitude is about
two times higher in the fully dynamical spacetime simulation (where
the BH is allowed to move).
We note that the growth rate and maximum $m=1$ mode amplitude in 
the simulation presented here is smaller than the one reported in
Paper I for the same model (the maximum of the $m=1$ mode amplitude 
is half of that reported in Paper I). The difference arises from the use of 
a constant damping parameter in the standard $\Gamma$-driver shift 
condition, which is known to be less accurate than dynamical 
damping parameters for large mass-ratio binary BH 
simulations (see for instance~\citet{Gold2013}).
The dynamics of the high mass-ratio quasi-binary system
composed of the central BH and the moving overdensity 
planet in the disc should therefore be described with 
better accuracy using the dynamical damping parameter
employed in Paper I. It seems that the fixed damping
parameter restricts the BH motion, resulting in a
smaller mode power. We note, however, that the actual
development of the instability and the overall properties
of the disc after its saturation remain unaffected by
the choice of gauge, as they should.

\subsubsection{Disk alignment}
\label{sec:alignment}

As shown by~\citet{King2005}, the direction of the torque
responsible for disc (and therefore also BH) precession
does not act in a direction to align the disc angular
momentum with the central BH. The alignment torque
ultimately results from the inclusion of the effects of
viscosity or dissipation, and its magnitude depends, 
as~\citet{King2005} remark, on the actual disc properties.
As our fluid evolution does not explicitly account for
viscosity, viscous and dissipative effects might therefore
only arise due to numerical viscosity and dissipation in 
shocks. As we have seen in Section~\ref{sec:spiral_density_wave},
the spiral density wave that results from the development of the
$m=1$ mode provides dissipation that is stronger than the one
arising from the numerics. 
Another factor working towards alignment of the BH spin and the
disc is the accretion of angular momentum, which is a negligible 
effect in our simulations due to the very low amount of total 
mass accreted during the evolution.

\begin{figure}
  \centering
  \includegraphics[scale=0.9]{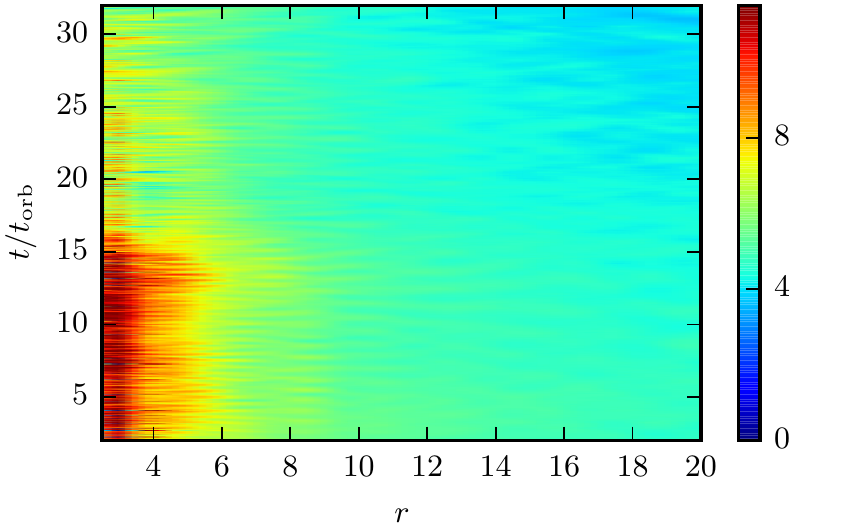}
  \caption{Spacetime diagram of the radial profile of the tilt angle $\nu(r)$. The tilt
    angle is shown in degrees.}
  \label{fig:tr_tilt}
\end{figure}

In Figure~\ref{fig:tr_tilt}, we plot a spacetime diagram showing the evolution of the 
tilt angle profile $\nu(r)$ obtained from the tracers. Similar to the 
disc analysis explained in~\citet{Nelson2000,Fragile2005} and 
employed in Paper I, we split the domain in radial shells and 
calculate the components 
of the total angular momentum of all tracers within a given shell
to obtain the tilt profile. 
We clearly see
the expected, non-zero oscillating tilt profile predicted
by the warp propagation as bending waves during the early stages
of the simulation. The tilt amplitude close to the central BH 
is significantly reduced from about $15 t_{\mathrm{orb}}$ onwards. This could be
connected to the time when the developing spiral density wave 
becomes non-linear for the first time, coupling to the fluid
inside the corotation radius and lowering the tilt amplitude.

\begin{figure}
  \centering
  \includegraphics[scale=0.9]{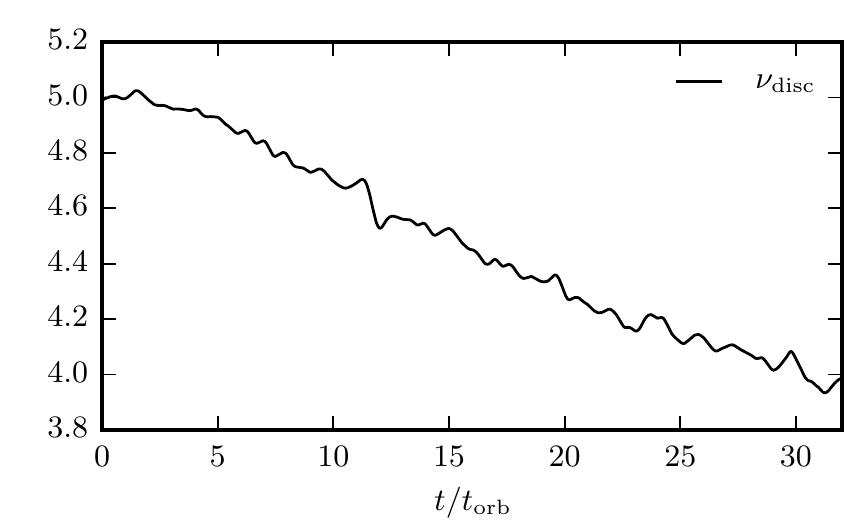}
  \caption{Time evolution of the global tilt angle of the disc, $\nu_{\mathrm{disc}}$, shown in degrees.}
  \label{fig:global_tilt}
\end{figure}

We do not see a complete global alignment as observed in~\citet{Kawaguchi2015}
during the time of the evolution. However, as
Figure~\ref{fig:global_tilt} shows, there is a clear monotonic drop
in the global tilt angle $\nu_{\mathrm{disc}}$ from $t=2\, t_{\mathrm{orb}}$ onwards,
which is when the $m=1$ non-axisymmetric mode starts growing.
\citet{Kawaguchi2015} noted that the 
alignment timescale was comparable or shorter than the precession
timescale of the disc. We note that our disc is precessing slower than the ones
studied in~\citet{Kawaguchi2015}, as our initial BH spin is much
smaller than that of the central BH they obtain in their BHNS merger
simulations. In our dynamical spacetime simulation, both the disc angular momentum and BH spin
have completed half a precession cycle within $32 \, t_{\mathrm{orb}}$,
while the global tilt angle $\nu_{\mathrm{disc}}$ has not dropped
by half within the same period.  A difference between our disc evolution and the 
ones described in~\citet{Kawaguchi2015} is the accretion timescale, which in their
case is comparable to the precession timescale and, therefore, to the alignment 
timescale. We do not observe this behaviour in our simulations, as the total rest 
mass accreted during the evolution ($2.6\times 10^{-3}$) is very small
compared to the initial disc rest mass ($1.02\times 10^{-1}$). Assuming that the 
accretion rate remains constant for the lifetime of the disc, the accretion timescale 
would be $\sim 10$ times larger than the precession timescale of the disc.

\subsection{QPOs in the accretion rate}
\label{sec:ecc_qpo}
\begin{figure}
  \centering
  \includegraphics[scale=0.9]{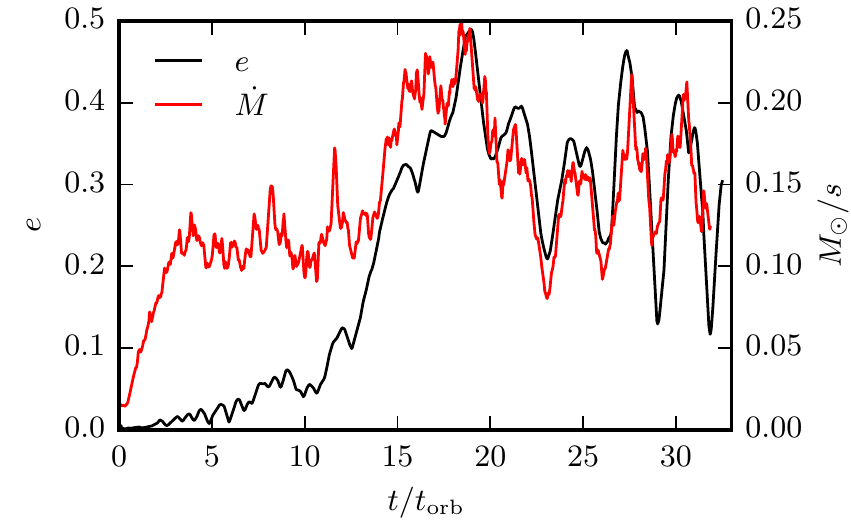}
  \caption{Evolution of the sum of disc eccentricity $e_1$ and 
disc ellipticity $e_2$ and of the rest mass
 accretion rate $\dot{M}$ in units of $M_{\odot}/s$. 
 The accretion rate has been shifted by $0.65 \, t_{\mathrm{orb}}$ in time for a better
 visualisation of the correlation between its modulation and that of $e$.}
  \label{fig:ecc}
\end{figure}

As already observed in Paper I, the growth of the $m=1$
non-axisymmetric mode in the disc strongly influences the
instantaneous accretion rate. As the accretion rate can be assumed to be 
a measure of X-ray luminosity (see, 
for instance~\citet{Paradijs1988,Mendez1999}), finding 
QPOs in the accretion rate and 
correlating them with the disc evolution might serve as
a model to explain the QPOs observed in LMXBs~(see e.g.~\cite{Klis2006} and references therein).
The origin of these QPOs is still not fully understood, 
and there are various models trying to explain the observed
X-ray variability in these sources (see~\citet{Lai2009} for a detailed
summary of the proposed models, ~\citet{Belloni2014} for a 
recent review of QPOs in LMXBs and~\citet{Mishra2015,Mazur2016}
for recent models).

\begin{figure}
  \centering
  \includegraphics[scale=0.9]{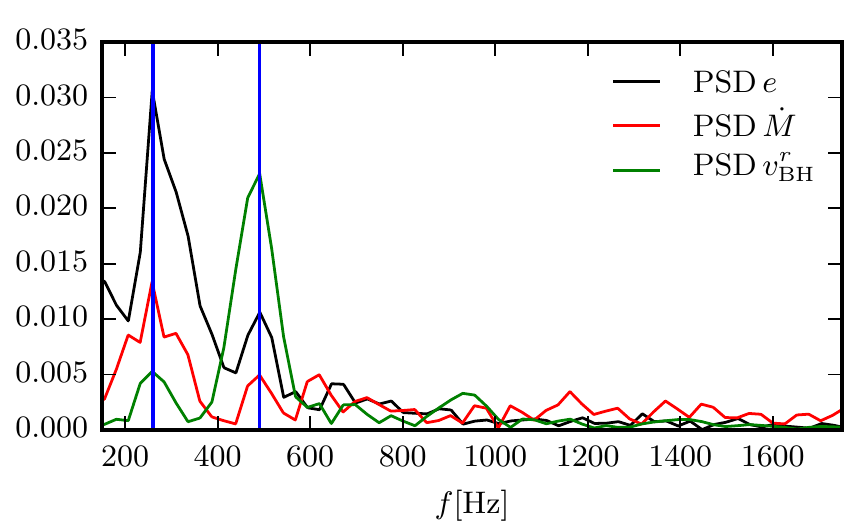}
  \caption{Power spectral densities of $e$, rest mass accretion
    rate $\dot{M}$, and radial BH velocity $v^r_{\mathrm{BH}}$. The vertical
    lines indicate the $260 \, \mathrm{Hz}$ and $490\, \mathrm{Hz}$ frequencies.}
  \label{fig:qpo}
\end{figure}

\begin{figure}
   \centering
   \includegraphics[scale=0.9]{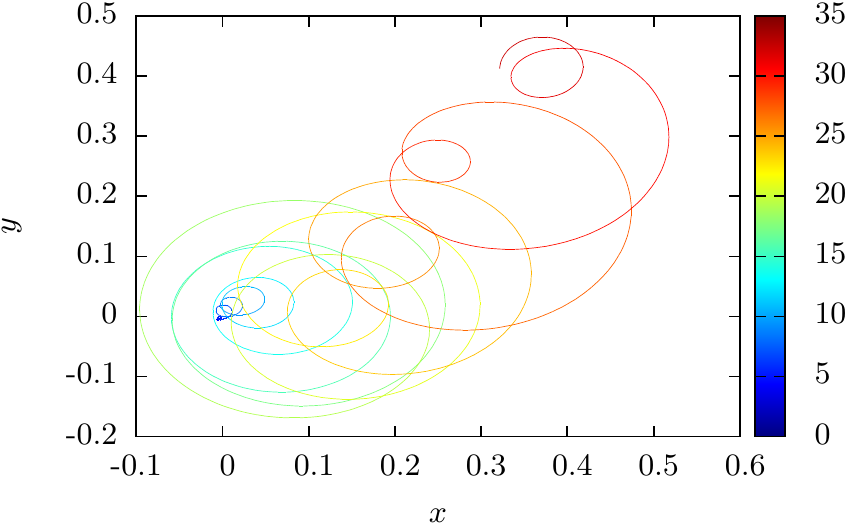}
   \caption{Trajectory of the BH projected onto the $xy$-plane. The
   colorbar indicates the time in $t_{\mathrm{orb}}$ along the trajectory. }
   \label{fig:BH_xy}
 \end{figure}

As matter accretion implies the development of a 
radial flow, we follow~\citet{MacFadyen2008} and  
calculate the eccentricity $e_1$ and ellipticity $e_2$ 
of the tracer particles as 
\begin{eqnarray}
e_n = \frac{|\sum_{j=1}^N m_j v^r e^{i n \phi}|}{\sum_{j=1}^N m_j v^{\phi}}\,,
\end{eqnarray}
where $v^r$ and $v^{\phi}$ are the radial and azimuthal components of
the 3-velocity, respectively. 
We denote the sum of disc eccentricity and ellipticity by
\begin{eqnarray}
e \equiv e_1+e_2.
\end{eqnarray}
The evolution of $e$ together with the mass accretion rate $\dot{M}$ is plotted in 
Figure~\ref{fig:ecc}. 
As in Paper I, we calculate the accretion rate as the following
surface integral at the AH:
\begin{eqnarray}
  \dot{M}= 2\pi r^2 \int_0^{2\pi} \int_0^{\pi} D \, v^r \, \sin \theta \, d\phi \, d\theta \,,
\end{eqnarray}
where $D\equiv \sqrt{\gamma}\rho W$ is the relativistic rest-mass density, and
$W$ is the Lorentz factor. The mass accretion rate in Fig.~\ref{fig:ecc} has been
shifted by $0.65 t_{\mathrm{orb}}$ to better illustrate that the accretion rate is clearly 
modulated by $e$ and shows distinct QPOs. As expected, the applied shift
is backwards in time, as the evolution of $\dot{M}$ trails the evolution of $e$.
The power spectral densities (PSD) of $e$, $\dot{M}$ and
the radial velocity of the BH are shown in Figure~\ref{fig:qpo}. The PSD
show a clear dominant peak at $\sim 260 \, \mathrm{Hz}$ with a first overtone
at $\sim 490\, \mathrm{Hz}$ for $e$ and $\dot{M}$. 
Note that the frequency of the first overtone is compatible 
with the orbital frequency of the $m=1$ pattern, which is $\sim 510$ Hz.
The frequency ratio of the dominant 
low frequency peak and the overtone is $o_1/f \sim 1.9$. The double peak in the PSD 
arises from the modulation of the strength of the $m=1$ mode (see Fig.~\ref{fig:PPI_growth_rate}).
The modulation causes changes in the eccentricity of the orbital motion of the BH as well, 
which is displayed in Fig.~\ref{fig:BH_xy}. 

Periodicity in the accretion rate due to the inner region of a circumbinary accretion
disc becoming eccentric has also been observed in~\citet{Farris2014}, as well as in the 
MHD simulations of~\citet{Machida2008}, where the authors attribute the QPOs in the 
accretion rate to the development of a $m=1$ non-axisymmetric mode. We note that the 
QPO frequencies extracted from our simulation are for a fiducial model with a central 
BH mass of $\sim 1\, {M}_{\odot}$ (see Table~\ref{tab:initial-model}). The observed 
frequencies in these systems usually scale as ${M}^{-1}$, where $M$ is the mass of the
central compact object, as in the $p$-mode torus oscillation model of~\citet{Rezzolla2003a}. 
Assuming the ${M}^{-1}$ frequency scaling to be universal (see for 
instance~\citet{Abramowicz2004,Zhou2015} for arguments and observational support), 
this would mean that the QPOs we observe in the accretion rate would be at $\sim 26$ Hz 
and $\sim 49$ Hz if we rescale our results for a $10\, {M}_{\odot}$ BH. These frequencies 
are compatible with the results of~\citet{Machida2008}, who found 10 Hz  QPOs for a 
$10\, {M}_{\odot}$ BH. Therefore, the QPOs we observe resulting from the modulation of 
the $m=1$ mode strength could help explain the low-frequency QPO sector of the fast X-ray 
variability seen in LMXBs. 

On the other hand, if we assume the accretor to be a NS, after rescaling our results for typical
NS masses,  the QPOs in the accretion rate would also have frequency peaks in a range 
compatible with those observed in LMXBs. Note that models based on the $p$-mode 
oscillations of axisymmetric tori cannot explain the observed twin QPOs in LMXBs with a 
NS as an accretor, with a fundamental frequency smaller than $500$ Hz, since the fundamental 
mode frequency decreases as the size of the disc increases or as the distribution of
the disc specific angular momentum approaches the Keplerian profile~\citep{Montero2012}.
Therefore, the $o_1/f$ frequency ratio in axisymmetric models shows a tendency to concentrate 
towards the $3:2$ ratio line as the fundamental mode frequency tends to zero. 
However, deviations from axisymmetry may relax this constraint, and there exist several works 
where the idea that the QPOs in accretion discs might be connected to non-axisymmetric modes 
has been put forward (see for instance~\citet{Li2003,Tagger2006,Machida2008,Lai2009,Henisey2012}).
By rescaling our results for a $1.4\, \mathrm{M}_{\odot}$ NS, the fundamental mode would lie 
at $\sim 185$ Hz while the first overtone would be at $\sim 350$ Hz. Such values, extracted from 
the modulation in the accretion rate triggered by the development of a non-axisymmetric $m=1$ 
instability, lie thus within the range needed to explain, for instance, the observed QPOs in sources 
like Cir X-1~\citep{Boutloukos2006} or XTE J1807-294~\citep{Linares2005}.

\section{Discussion}
\label{sec:conclusions}

We have presented results from three-dimensional, numerical relativity simulations 
of a {\it tilted} black hole-thick accretion disc system. In particular, we have investigated in detail 
the BH--torus dynamics of one specific model, {\tt NC1a03b05}, from the large set of models discussed
in~\citet{Mewes2016}. In this paper we have employed fluid tracer particles
as a powerful new tool to analyse the disc dynamics. The tracers provide a complementary tool 
to the existing disc analysis thorn described in Paper I.

Using the tracers, which accurately describe the bulk morphology
of the fluid in the disc, we have shown that the BH precession we 
reported in Paper I is indeed caused by a 
torque resulting from the disc precessing as a solid body, which
in turn results from the LT torque the BH exerts on the disc.
This is expected, as the disc should exert an equal and
opposite torque on the BH. For sufficiently 
high disc-to-BH mass ratios, tilted BH--torus systems should therefore 
contain precessing central BHs, and the BHs should precess for
at least the accretion timescale of the disc.

The main characteristic of the BH-torus model considered is its non-constant
specific angular momentum profile. As already observed in the numerical relativity 
simulations of~\citet{Kiuchi2011} and in Paper I,
these models form a long-lived over-density \enquote{planet}
as the result of the growth of the PPI. Using the tracers, 
we have shown that the growth of the $m=1$ non-axisymmetric
instability manifests itself as a spiral density wave of
constant pattern speed in the differentially rotating
disc. The pattern period has been shown to be slightly 
smaller than the initial fluid period at the 
location of the maximum rest-mass density of the disc. While
the spiral wave remains of small amplitude, it travels
through the fluid without interaction. Once the wave amplitude 
becomes large enough for non-linear effects to become 
important (in some regions the fractional change in
rest-mass density becomes as high as 0.9), the wave couples 
to the fluid via the formation of mild shocks, which can be 
seen clearly as an increase in the fluid entropy. 
As the disc is differentially rotating, the spiral density wave has 
{\it negative} angular momentum w.r.t. the fluid inside the 
so-called corotation radius; therefore, the wave can 
transport fluid angular momentum outwards and becomes the main
driver for accretion. 

The density wave also represents
a time-changing non-axisymmetric gravitational potential.
As observed in~\citet{Korobkin2011} and 
Paper I, this gravitational potential causes 
the BH to move along a spiral trajectory. Using the tracers,
we have calculated the total gravitational pull exerted by
the disc on the BH and have shown that the resulting torque
agrees very well with the observed motion of the BH. This 
is an important confirmation that the BH motion is of physical
origin, and not caused by numerical effects or the evolution
of the gauge variables lapse and shift.
BH--disc alignment within the accretion timescale was
observed in the numerical relativity simulations of tilted post-merger discs
in~\citet{Kawaguchi2015}. Correspondingly, in the simulations 
reported in Paper I and in this work
we have also seen partial realignment of the BH--torus system. In all
cases, there was no explicit viscosity in the fluid evolution.
As the torque acting to align the BH spin with the disc must
be of dissipative nature, the only way to achieve BH--torus
alignment is via numerical dissipation or dissipation via 
shocks in the fluid. The spiral density wave that results
from the growth of the $m=1$ mode provides such a channel, 
and~\citet{Kawaguchi2015} have speculated that
shocks in non-axisymmetric waves might be responsible
for the observed alignment of the system. This is a strong argument 
to carefully check for non-axisymmetric structures in 
post-merger disks, as they seem to influence the evolution of 
BH-torus systems significantly.

Our simulations have also revealed the presence of distinct QPOs 
in the evolution of the accretion rate, in a frequency range compatible
with that of X-ray luminosity QPOs in LMXBs. When rescaling
the frequency of the observed QPOs in our simulation for
a $10 {M}_{\odot}$ BH, the extracted frequencies
are compatible with the range of low-frequency QPOs in those
systems. Furthermore, the same rescaling for typical NS masses 
also gives QPOs with frequencies compatible with those observed in sources 
like Cir X-1 or XTE J1807-294. The frequency ratio of the dominant low frequency 
peak and the first overtone found in our three-dimensional simulations is $o1/f \sim 1.9$, 
a frequency ratio not attainable when modelling the QPOs as $p$-mode oscillations 
in axisymmetric tori. As the flow needs to develop a
non-zero radial velocity component in order to accrete,
we have also analysed the sum of disc eccentricity and
ellipticity using the tracers and we have shown that its evolution
exhibits the same QPO structure at exactly the same frequencies.
 While the origin of the variability of the eccentricity (which is also exhibited in the
radial motion of the BH) is still unclear, the fact that the accretion
rate is so clearly modulated 
could be the starting point to devise a new model to explain the
observed QPOs in LMXBs.

\section*{Acknowledgements}

 Research supported  
  by the Spanish Ministry of Economy and Competitiveness (MINECO) through grant 
  AYA2013-40979-P, by the Generalitat Valenciana  
  (PROMETEOII-2014-069), by the Max Planck Institute for Astrophysics, and by the Helmholtz 
  International Center for FAIR within the framework of the LOEWE program launched by the 
  State of Hesse. FG would like to thank David Radice for the useful discussions and for the 
  help with the implementation of the tracer particles.

\bibliographystyle{mnras} \bibliography{references}

\end{document}